# REVIEW OF SYN-FLOODING ATTACK DETECTION MECHANISM

Mehdi Ebady Manna and Angela Amphawan

School of computing, University Utara Malaysia, Kedah, Malaysia
meh_man12@yahoo.com; angela@uum.edu.my

## ABSTRACT

*Denial of Service (DoS) is a security threat which compromises the confidentiality of information stored in Local Area Networks (LANs) due to unauthorized access by spoofed IP addresses. SYN Flooding is a type of DoS which is harmful to network as the flooding of packets may delay other users from accessing the server and in severe cases, the server may need to be shut down, wasting valuable resources, especially in critical real-time services such as in e-commerce and the medical field. The objective of this paper is to review the state-of-the art of detection mechanisms for SYN flooding. The detection schemes for SYN Flooding attacks have been classified broadly into three categories – detection schemes based on the router data structure, detection schemes based on statistical analysis of the packet flow and detection schemes based on artificial intelligence. The advantages and disadvantages for various detection schemes under each category have been critically examined. The performance measures of the categories have also been compared.*

## 1. INTRODUCTION

Information has become an organization's most precious asset. Organizations have become increasingly dependent on information. The widespread use of e-commerce has increased the necessity of protecting the system to a very high extent[1]. Denial of Service (DoS) is a security threat which compromises the confidentiality of information stored in Local Area Networks (LANs) due to unauthorized access by spoofed IP addresses. DoS is harmful to LANs as the flooding of packets may delay other users from accessing the server and in severe cases, the server may need to be shut down, wasting valuable resources, especially in critical real-time services such as in e-commerce.

The term security refers to the protection of essential information against unauthorized access. There are many types of threats that violate the information that can be accessed by global network: theft and fraud, hackers, Denial of Service (DoS) and virus [2](see Table 1) below. In addition, it is difficult to make a global secure system for the entire network (i.e. the Internet).

Table 1: Examples of Threats

| Threat | Theft and Fraud | Loss of Confidential | Loss of Privacy | Loss of integrity | Loss of Availability |
|---|---|---|---|---|---|
| Using another person's means of access | √ | √ | √ | | |
| Unauthorized amendment or copying of data | √ | | | √ | |
| Program alteration | √ | | | √ | √ |
| Illegal entry by hacker | √ | √ | √ | | |
| Blackmail | √ | √ | √ | | |
| Thefts of data , programs and equipment | √ | √ | √ | | √ |
| Inadequate staff training | | √ | √ | √ | √ |
| Viewing and disclosing unauthorized data | √ | √ | √ | | |
| Introduction of virus | | | | √ | √ |

Any violation in security leading to loss of confidence and loss of privacy could lead to illegal action taken against an organization. On the other hand, loss of data integrity results in invalid or corrupted for information. Hence, losing any factor of these criteria would cause significant harm in business for the organization assets such as loss customer confidence, contract damages, regulatory fines and restrictions, or a reduction in market reputation. In the worst case, a failure to control or protect information could lead to significant financial losses or regulatory restrictions on the ability to conduct business.

## 2. NETWORK THREATS

A network threat is any form of security breach which is intended for access to protected information or corrupt it [3]. For example, "viewing and disclosing unauthorized data is a threat which may result in theft and fraud, loss of confidentially and loss of privacy for the company. In addition, the important activities are continuously running and users remain unaware of known vulnerabilities and the patches that address them. To this end, a useful security tool is to identify how a particular application advises users of the existence of security vulnerabilities and availability of patches to address them (by newsgroup posting, mailing list, corporate web site, etc.) [4]. Network threats may include any of the following:

**a.** Wiretapping: intercepting the communications by a passive or active way.
**b.** Impersonation: pretending to be someone (personnel) or something (process).
**c.** Hacking:breaking down the computer system security.
**d.** Virus, Trojans and worms, the definitions as below:
  - A computer virus is a hidden program written to alter the way the computer operates. It attaches itself to a program or file enabling it to spread from one computer to another and causing infections as it is travels.
  - Worms are a sub-class of virus in design and code written. It has the ability to spread from computer to computer, but unlike a virus, it has the capability to travel without any human action.
  - Trojan is a program that appears to be useful software but will actually do damage once installed or run on computer. For example, Trojan horse.
**e.** Denial of Service Attacks (DoS).

## 3. THE CONCEPT OF DENIAL OF SERVICE (DoS) ATTACK

In order to understand previous forms of detection for Denial of Service (DoS) attacks, it is important to first understand the concept of DoS attacks. There are several types of DoS attacks. In this paper, the focus is on a type of DoScalled SYN-flooding.  An attacker in SYN-flooding sends a succession of SYN request to a victim's system in an attempt to make a system unresponsive to legitimate traffic. The SYN-flooding attacks exploit the TCP's three-way handshake mechanism and its limitation in maintaining half-open connections. The SYN-flooding attacks' abuse network resources and can bring about serious threats to the internet. On the other hand, the SYN-flooding attack is networks anomalies usually refer to the conditions when network operations diverge from the normal behavior [5]. Anomaly detection in a network is a very complex task for SYN-flooding attack, because it is dependent upon the nature of the data that is available for the analysis.

The SYN-flooding data can be classified into two major types: the first is network-based and the second is end-user-based [6], [7].

This attack can cause significant financial losses in the client/ server network, especially in e-commerce. When a server receives a SYN request, it returns a SYN/ACK packet to the client. Until the SYN/ACK packet is acknowledged by the client, the connection remains in a half-open state for a period of up to the TCP connection timeout, which is typically set to 7.5 seconds during the SYN-flooding attacking time[8]. The half-open connection keeps the server in waiting state to be acknowledged by the client. However, the server has built in its system memory a backlog queue to maintain all half-open connections. Since this backlog queue is of finite size, once the backlog queue limit is reached to the maximum limited number, all connection requests will be dropped within SYN-flooding attack's time  [9], [10].

## 4. TYPES OF DENIAL OF SERVICE (DoS)

Denial of Service (DoS) attack is a common threat from the user which disrupts a service on the network. It can be defined as any action or series of actions that prevent any part of a telecommunications system from functioning. A very common example is the SYN-flood attack, which is simply to send a large number of SYN packets and never acknowledge any of the replies and bombarding the service with such a large number of pointless requests that the serious users are unable to use it. See (Fig.1).

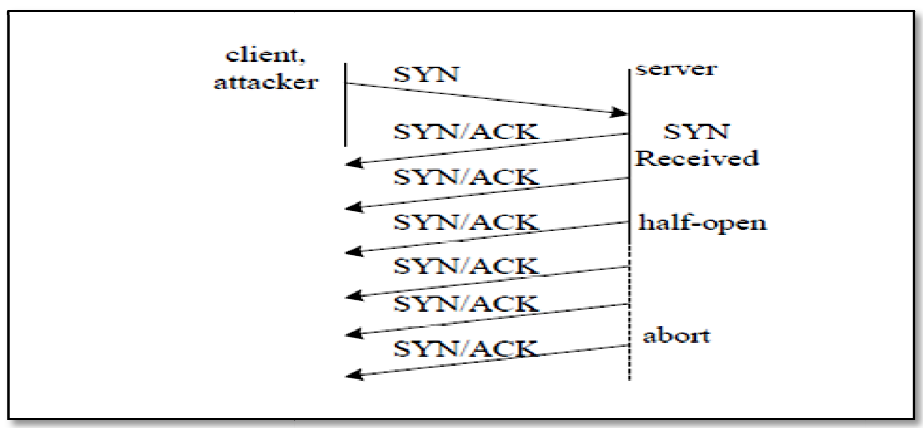

Figure 1. SYN-Flooding attack scenario

Some other examples of DoS types include Ping Flood and Ping of Death which less on larger networks or websites because only one computer is being used to flood the victim's resources as below :

- **Ping flood:** It relies on the ICMP echo command, more popularly known as ping. It is the most basic of attacks, which used by network administrators to test connectivity between two computers. In the attack case, it is used to flood large amounts of data packets to the victim's computer in an attempt to overload it.

- **Ping of Death:** it can cripple network resources based on a flaw in the TCP/ IP. The maximum size for a packet is 65, 535 bytes. If one sender were to send a packet larger than that, the receiving computer would ultimately crash from confusion.

    It happens when the hackers can bypass this by cleverly sending the packets in fragments. When the fragments are assembled on the receiving computer, the overall packet size is too great, which cause the buffer overflow and crash the

-

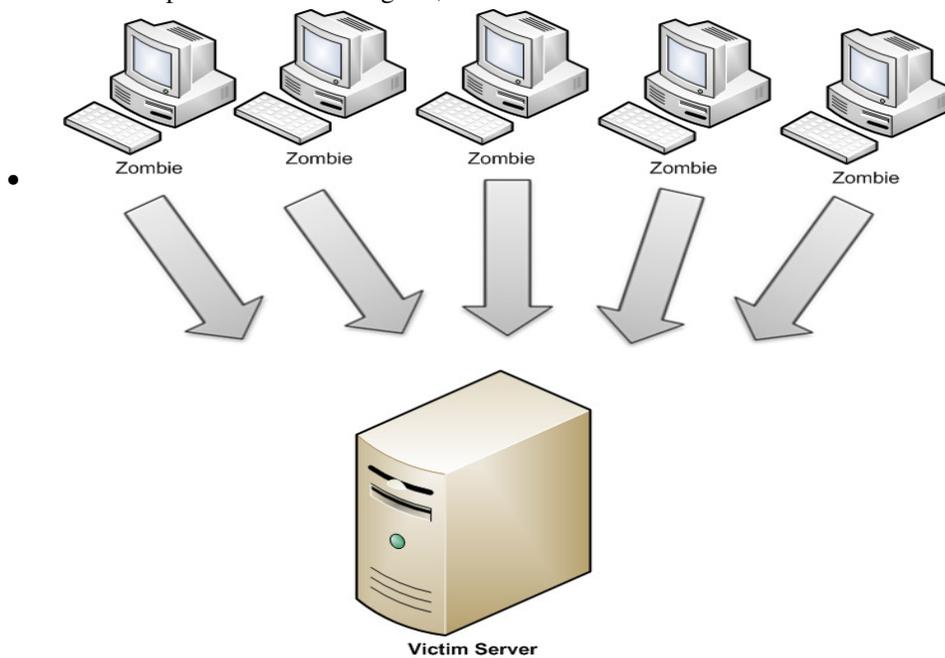

- **ial of Service (DDoS):** this kind of attack is considered the worst one in denial of service attacks where multiple computers are being used. In this instance, the computers that are being used may or may not be aware of the fact that they are attacking a website or network. Trojans and viruses commonly give the attacker control of a computer. The victims of computers are called zombies. See (Fig. 2).

Figure 2. Distributed Denial of Service (DDoS) Attack

## 5. IP- SPOOFING BASED TCP/IP

IP spoofing refers to the transmission of IP packets with forged IP addresses to the victim computer. IP spoofing is most used in the internet in denial-of-service attacks. The victim computer will be flooded with useless information which overwhelms the network resources. The attacker does not wait to receive responses from the receiver of the attacked packet. The packets of this type of attack are difficult to filter since each packet appears to come from different addresses, and the source IP address of the attack is hidden. On the other hand, the attacker aims to flood the server with such a useless packet causing the server to be shut down [11].

If a SYN request is spoofed, the victim server will never receive the final ACK packet to complete the three-way handshake.

Flooding spoofed SYN requests can easily exhaust the victim server's backlog queue, causing all the incoming SYN requests to be dropped. The stateless and destination-based nature of the Internet routing infrastructure cannot differentiate a legitimate SYN from a spoofed one, and TCP does not offer strong authentication on SYN packets [12][13].

Under IP spoofing, however, the three-way handshake will be very different from that of the normal case. Attackers usually use unreachable spoofed source IPs in the attacking packets to improve the attack efficiency [14].

# 6. PREVIOUS WORK ON DoS METHODS

Many of the information resources that are made available and maintained inthe Internet have a high intrinsic value to their users. Their security is very important. As mentioned above, security of information resources has three components: confidentiality (protection against disclosure to unauthorized individuals); integrity (protection against alteration or corruption); and availability (protection against interference with the means to access the recourses).
Security risks are associated with allowing free access to all of the resources in an intranet or in a local area network. Many organizations adopt a firewall to make a barrier around an intranet. Nevertheless, networks are too complex to be defended using only the traditional shielding techniques of cryptography, authentication and static firewalls. Most of the security application (e.g. firewall) devices are unlikely to be able to prevent (DoS) attacks for many reasons:

    A. The attacking traffic is likely to closely resemble real service requests or responses.
    B. Even if they can be recognized as malicious, a successful attack is likely to produce malicious messages in such large quantities that the firewall itself is likely to be overwhelmed and become a bottleneck, preventing communication with the services that it protects [15].

Many methods have been proposed to detect the denial of SYN-Flooding attacks. A hash table is a high-performance method to detect such attack. [16] adopted a technique based on the hash table for IP traceback, which generates audit trails for the network trace so that the origin of an IP packet delivered by the network in the recent past can be traced.

A hash table has also been employed to look for imbalance between the incoming and outgoing trace packet flows to or from an IP address [17]. The flow imbalance can facilitate the detection of DoS. In normal packet flow, the number of incoming packet is matched to the number of outgoing packets over a period of time. For example, each packet in TCP connection is normally acknowledged. However, during the attacking time, the numbers of incoming and outgoing packets are imbalanced. In the case of attacking source, the number of outgoing packets exceeds the number of incoming packets. This imbalance can be used to infer that the attack is occurring. The following schemes have been presented for detecting of SYN-flooding attacks.

# 7. CLASSIFICATION OF DoS DETECTION SCHEMES

DoS detection schemes may be classified into three categories as shown in Fig.3:

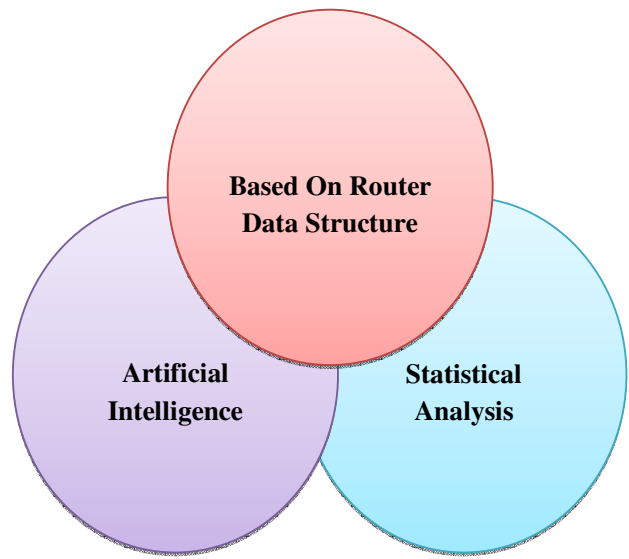

Figure 3. Denial of Service Attacks (DoS) classification schemes

7.1 Based on router data structure.
7.2 Based on statistical analysis for packet flow.
7.3 Based on Artificial Intelligence (fuzzy logic and neural network).

## 7.1 Router-Based Detection Scheme using Bloom-Filter

A bloom filter is a space-efficient data structure used in router for pattern matching in many network communications. It is used to inspect packets and detect malicious packets based on many algorithms [18],[19].

[20]focuses on the low-rate agent and present a router-based detection scheme for it. The low-rate DoS agent exploits the TCP's slow time scale of Retransmission Time Out (RTO) to reduce TCP throughput. In this case, the DoS attacker can cause a TCP flow in RTO state by sending high rate requests for short-duration bursts. Therefore, The TCP throughput at the victim side will be reduced during the attacking time on low-rate DoS agent. The proposed scheme is based on the TCP SYN-SYN/ACK protocol pairs with the consideration of packet header information (both sequence and Ack. Numbers). The Counting Bloom Filter (CBF) is used to avoid the effect of ACK retransmission, and the change point detection method is applied to avoid the dependence of detection on sites and access patterns. See (Fig.4).

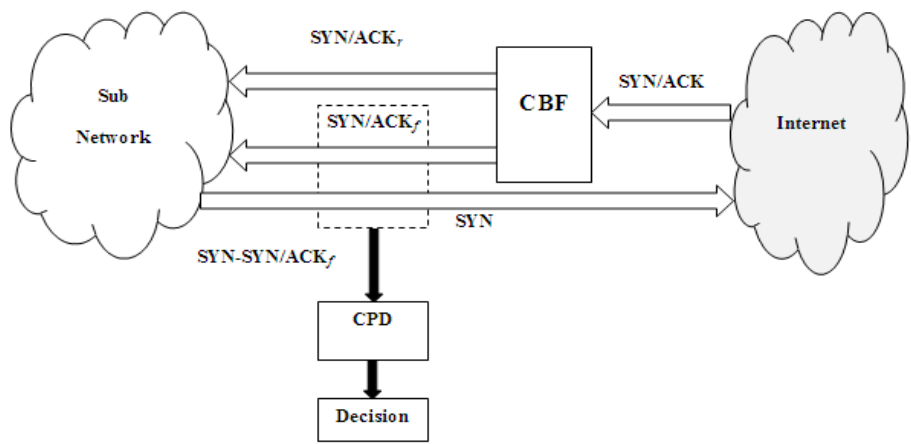

Figure 4: Router based Counter Bloom Filter (CBF) Scheme

Tracebackbased Bloom Filter (TBF) was adopted to record the TCP session statistics Internet Protocol Time To Live (IP-TTL) of SYN packets. As the attacks start, the SYN packets and IP-TTL statistics were matched to differentiate the attacks' packets and record IP- TTL, because SYN-flooding attacks, are too time-expensive and consume the resources such as the memory. For instance, when SYN-flooding started, victim servers have to call for a lot of memory, usually more than 500MB, to store the attack packets[21].

The main advantages and disadvantages for each scheme are summarized in Tables below. For the router based detection scheme, the main advantages and disadvantages of important papers are critically examined in Table 2.

Table 2: Advantages and Disadvantages for router Scheme

| paper | Technique | Disadvantages | Advantages |
|-------|-----------|---------------|------------|
| [22] | This technique adopted SYN flooding detection based on Bloom filter (SFD-BF) in an edge router to detect SYN flood attack by recording packet information for TCP-FIN pairs behaviour. | - False Positive (FP) is generated due to bloom filter data structure.<br>- This technique is inefficient if in case of using (FIN) in the next SYN packet. | - The main advantages of this technique is it used the Change Point Detection method based on non-parametric Cumulative Sum (CUSUM) to avoid the discrepancy in small number of long lived TCP and retransmission of SYN packets. |
| [20] | This technique focuses on low-rate agent, and the proposed system scheme is based on TCP | All incoming SYN/ ACK packets were classified based on Counter Bloom Filter (CBF) data | Uses the counting bloom filter data structure to detect the behaviour of TCP SYN- SYN/ ACK, |
| | SYN – SYN/ ACK protocol pair with consideration of packet header information. | structure for the sub networks. If there is multiple packets flow, a CBF can increment and cause *false results* in the classification phase. | and the new scheme shows a shorter time to both low-rate and high-rate attacks. |
| [21] | Records the TCP session's statistics (IP-TTL) of SYN packets in a trace back- based bloom filter. Then match the SYN packets and IP-TTL statistics to differentiate the attack's packets. | False positive (FP) results have been seen when using bloom filter data structure in multiple packet's case. The false positive occurs when the detection method mistakenly flags a normal traffic as being attacked and this way causing the difficulty to measure the detection accuracy. | - Uses statistical measurement to record TTL from IP trace back into 256 areas of trace-bloom filter data structure. <br> - In addition, it shows efficient results when packets are sent based on these groups. |
| [23] | This technique presented of SYN/ACK- CliACK (SACK) pair in leaf router to identify outgoing SYN/ACK in the victim server and the TCP CliACK port that being attacked. The SYN/ACK packet is sent by the server and CliACK packet is ACK packet sent by the client to complete the three-way handshake. | - This technique is generated a False Positive (FP) values due to the use of bloom filter data structure in the case of congestion status in network flow <br> - This technique is inefficient in the case of Request- Reply (RR) protocol which sends the CliACK in the next SYN request. | - This technique presented a quantitative analysis results based on Counting Bloom Filter (CBF) in leaf router. The CBF is used to store the full information for TCP connection which including client and server IP address, ports and initial sequence numbers. <br> - Low consumption of memory resources. The memory cost of SACK technique is 364 K.B for 10GBS link flow in the worst case. |
| [24] | Detecting the validity of outgoing SYN and incoming SYN/ACK in edge router is proposed. The edge router connects end hosts to the internet. The mapping process is accomplished in a hash map table by checking pairs of outgoing SYN request and incoming SYN/ACK that store in data base. | - Integration process of storing the packet information such as source and destination IP addresses is difficult in the case of congestion in the network flow. This leads to incorrect values in the mapping table of bloom filter data structure. | - This technique guarantees that each packet sent by the client is valid by two parts which storage module and inspection module. <br> - The hash table is efficient to store source and destination IP address in the database at the storage module. <br> - This technique shows an accurate detection SYN flood or abnormal case attack based on mapping table which is |

| | | | constructed in the storage module. |
|---|---|---|---|
| [25] | This paper is used to examine the flooding attacks against VoIP architectures that employ the Session Initiation Protocol (SIP) based on bloom filter data structure in router. The monitor process based on bloom filter combined with a new metric " session distance" which should equal zero in normal operation | - This technique is mainly increased the resource consumption such as CPU time and memory and bandwidth during the DDoS flood attack because it is established sessions between the requester and responder.<br>- It is generated False Positive (FP) and False negative (FN) values due to data structure bloom filter. | - This approach is successfully reflected the behaviour of detection of certain SYN flooding attacks in Session Initiation Protocol (SIP) based VOIP and inefficient to launches against other signaling protocol H.323 and MCGP. |

## 7.2 Sample Flow-Statistical Analysis

Many efforts have been undertaken in using the sample flow of statistics to detect DoS attacks [26]-[28].

[29]presented a statistical scheme to detect the SYN-flooding accuracy on network anomalies using flow statistics obtained through packet sampling. The network anomalies generate huge number of small flows, such as network scans or SYN-flooding. Due to this reason, it is hard to detect SYN-flooding when performing packet sampling because the network flow may be either bursty (non-linear) or under the normal flow rate. Their model is based on two steps: the first step, analytical model was developed to quantitatively evaluate the effect of packet sampling on the detection accuracy and then investigated why detection accuracy worsens when the packet sampling rate decreases. In addition, it is shown that, even with a low sampling rate, the detection accuracy was increased because the monitored traffic was partitioned into groups. The results show that the proposed mechanism is demonstrated to have the capability of detecting SYN-flooding attack accurately.

According to [30], a new detection method for DoS attack traffic based on the statistical test has been adopted. Investigation of the statistics of the SYN arrival rate revealed that the SYN arrival rate can be modeled by a normal distribution. A threshold for maximum arrival rate to detect SYN-flooding traffic has been established. In addition, the threshold for incomplete three-way handshaking packet ratio to detect possible DoS traffic also has been determined. This mechanism was shown to be effective in detecting SYN-flooding attack, but for the normal traffic threshold, the value is not accurate for the whole packet flow, especially during the attacking time.

For statistical analysis, the main advantages and disadvantages of important papers are critically examined in Table 3.

Table 3: Advantages and Disadvantages for statistical analysis

| paper | Technique | Disadvantages | Advantages |
|-------|-----------|---------------|------------|
| [29] | An analytical model was developed to quantitatively evaluate the effect of packet sampling on the detection accuracy. | - The proposed analytical model is not able to detect SYN-flooding when the sampling rate is low.<br>- In addition, the statistical analysis limits the performance of network communication because of the overhead for sampling packets in real time. | Uses a threshold value to detect the anomalies in the flow rate which is determined based on normal traffic statistics. In addition, the traffic has been partitioned into groups to increase the detection accuracy of network anomalies where the low-rate anomalies have been detected based on normal traffic statistics. |
| [30] | The proposed technique is based on the statistical test for SYN arrival rate. In addition, many simulation experiments were used to determine the threshold value for traffic flow in normal distribution. This threshold value has been checked with SYN-flooding threshold value to detect such attack. | - There is no explicit threshold value for TCP SYN/ SYN-ACK to determine for the normal traffic and difficult to determine this value in low rate because of false positive and false-negative results.<br>- In addition, modelling and estimation network traffic is difficult because the network traffic has linear and burst characteristics. | - The main advantage of this scheme is to control the statistical measurement under the threshold value in normal distribution of simulation experiments.<br>- In addition, it reflects the behaviour of packet's flow and how the false positive and false-negative values can be measured. Furthermore, the required parameters for the threshold value has been determined in normal distribution. |
| [31] | Abnormal statistics method based on correlation analysis is developed in this technique. The method is based on extracting anomalies space from network wide traffic in | - This technique is generated very high False Positive (FP) results.<br>- This technique might not be effective if the attacker is cleverly | - This technique is successfully detected the DDoS by monitors the derivation in correlation analysis across network traffic from normal |
| | the backbone network. | send attack under certain leave (low rate attack) | case.<br>- This technique achieves good results of detecting small intensity attacks compared with other methods such as Principle Component Analysis (PCA). |
| [32] | Statistics method based on mean to detect the SYN flood attack is developed in this technique. The matching process is conducted by comparing the difference between the overall means analysis of incoming traffic arrival rate and normal traffic arrival rate. | - This technique cannot overcome the low-rate SYN flooding attack which happens on condition that the arrival rate difference between attacks and normal.<br>- False Negative (FN) and False Positive (FP) values are generated in this technique.<br>- The technique is mainly included resources consumption in case of low rate attacks which leads to shut down the available resources. | - The main advantage of this technique is low computational overhead because the proposed scheme does not hold the three-way handshake states but only statistically analysis the SYN and ACK segments. |
| [33] | An Agent-based Intrusion Detection System (AIDS) is proposed based on Chi-Square statistical method to detect Dos/DDoS. | - This method is based on statistics analysis and does not reflect the behaviour and reliability of agent-based Intrusion Detection System (AIDS).<br>- The statistical analysis limits the performance of the communication because of the overhead in sending packets.<br>- False Positive (FP) values are generated based on this method. | -The main advantages of this method is statistically analyse amount and variation of packet issued by the sender. The first step in this method is to find out the Chi-Square and after that check whether or not it exceeded the threshold of the normal distribution. |
| [44] | Originalmathematical | - The statistical | - In this scheme, the |
| | measurements approach based on entropy to detect SYN-flooding attacks from the victim's side is proposed in this technique. The technique is monitored unusual handshake sequences to calculate the entropy. This value is used to detect the changes in the balance of TCP handshakes. | analysis limits the performance of the communication because of the overhead.<br>- False positive (FP) was generated in this technique. | accuracy detection is done in real-time to allow quick protection and help guarantee a proper defense.<br>- Experimental results show that the method can detect SYN-flooding attacks with better accuracy and robustness methods |
| [45] | - A real-time DDoS attack detection and prevention system deployed at the leaf router to monitor and detect DDoS attacks was proposed.<br><br>- The leaf router respectively real timely analyzes and detects the clients' and servers' traffic in the case of SYN flooding attacks and normal operations.<br><br>- This system periodically samples the traffic flow from every IP address and judges whether its traffic behavior meets the synchronization or not. | - CPU time and memory consumption overhead is seen in this technique. | - The main advantages in this technique:<br>- The system can recognize attackers, victims and normal users, and filter or forward IP packets by a quick identification technique.<br>- The identification technique was checked the mismatch between the request packets and response packets.<br>- The experiment results show that the system can make a real-time detection for flooding attacks at the early attacking stage, and take effective measures to quench it. |

## 7.3 Detection Scheme using Fuzzy Logic and Neural Nets

Fuzzy logic and neural network was adopted by many research to design and implement intrusion detection systems for denial of service attacks[34]-[36].

A fuzzy logic based system for detecting SYN-flooding attacks has been adopted. Fuzzy logic helps solving the systems which have elements of uncertainly. Fuzzy logic is appropriate for approaching the nonlinear systems [37].

[38] proposed a system represented by two blocks shown in figure 5. The first one is the packet classification block which classifies incoming network traffic packets, where the header of each captured packet is checked to see if it is a TCP SYN packet; if the fragment offset value in the header is zero, then it is a TCP packet. If the SYN flag of the flag bits in this TCP packet is one, then it is a SYN packet (attack possibility). The packet classification block collects the TCP SYN packets for a predetermined Δt time and gives them to the fuzzy logic system, which is the second block of the proposed system. The Δt in this work was 5 seconds, while the second block of the proposed system is a fuzzy logic system. This block is responsible for SYN-flooding attack detection. The detection accuracy of the proposed system was compared with Cumulative Sum (CUSUM) for five attacks and showed a high accuracy and low false-negative rate and generate an earlier alarm than CUSUM algorithm which it is an ideal algorithm for identifying DoS attacks based on the measurement for the mean in traffic before, and after they detect comparing with the threshold value.

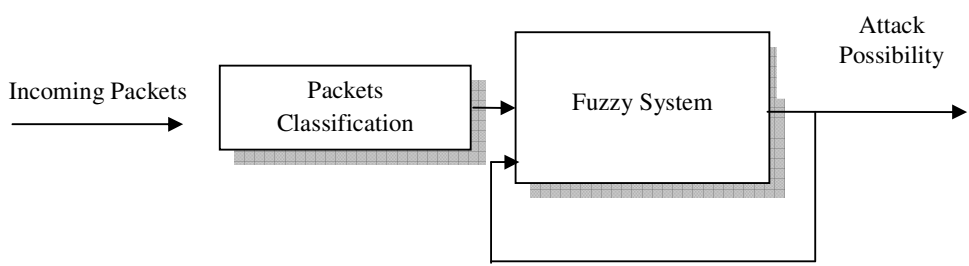

Figure 5: The proposed system based on fuzzy logic [38]

On the other hand, [39], adopted a hierarchical off-line anomaly network intrusion detection system based on Distributed Time-Delay Artificial Neural Network(DTDANN). The work aims to solve a hierarchical multi class problem in which the type of attack (DoS) detected by a dynamic neural network. Actually, the data set used by this method is old data and does not reflect the current behavior of attack packets.

For fuzzy logic and neural network detection techniques under the carter gory of artificial intelligence, the main advantages and disadvantages ofimportant papers are critically examined in Table 4:

Table 4: Advantages and Disadvantages for DoS detection based on artificial intelligence

| paper | Technique | Disadvantages | Advantages |
|-------|-----------|---------------|------------|
| [40] | A hybrid Intrusion Detection System based on Artificial Neural Network Self-Organizing Maps (SOM) and Resilient Propagation Neural Network (RPROP) is proposed to detect to visualize and detect SYN flooding malicious. | - Old off-line data set is used from the third international Knowledge Discovery and Data Mining Tools Competition (KDD Cup) since 1999 which is not reflect the current behaviour of modern SYN flooding attacks. <br> - Training phase for both neural network are used based on | - Quantitative and qualitative analysis has been performed by using hybrid intrusion detection system based on SOM and (RPROP). |

| | | | deterministic features from the same old dataset | |
|---|---|---|---|---|
| [38] | The technique uses the fuzzy logic for nonlinear systems to detect DoS attack. | - | It is difficult to model the traffic network before, and after the attack due to linear and burst characteristics of packets flow. <br> - In addition, this technique depends on offset value in a TCP packet header which is a change due to network congestion and others states. | This technique shows a good result using a fuzzy logic compared to CUSUM algorithm, especially in false positive and false-negative measurement in the low-rate flow packets. |
| [39] | This technique adopts a hierarchical off-line anomaly network detection system based on distributed time delay artificial neural network. | - | The data set of this scheme has been selected from DARPA intrusion detection data set in 1998 which is does not reflect the traffic model in the recent traffic, especially the new denial of service type's attacks. <br> - In addition, most neural network requires retraining to improve analysis on varying input data, and it is impossible in an on-line traffic flow. <br> - In this scheme also, off-line experiments have been conducted in neural network, which is associated with the interconnection among processing elements and set of desired output and transfer between input and output is determined by weights. | The main advantages is to build off-line detecting system based on the distribution time delay artificial neural network to detect multi class of denial of services (DoS) attacks. |
| [41] | Alert Classification System is developed by | - | Redundant alert is triggered for the same | - Many DDoS flood attacks such as SYN flood TCP, |

| | using Machine Learning Algorithms Artificial Neural Networks and Support Vector Machines in this technique. | attack attributes.<br>- This technique includes overhead during the training phase of neural network and cannot integrate with new type of attacks. | UDP and ICMP have been detected in this based on neural networks technique.<br>- This technique helps to generate many DDoS attacks types from real-time independent network which can help in generalization and reflecting the nature of the DDoS attacks. |
|---|---|---|---|
| [42] | Identification the occurrence of DDoS intensity in real time based on Fuzzy logic is developed in this technique | - Light or moderate intensity flood attack is missed during the classification phase of Fuzzy logic which can cause more False Negative (FN) values.<br>- Difficult to set Fuzzy logic during DDoS attack because the attacker sends in nonlinear way.<br>- Light (low – rate)DDoS flood attack is not detected properly in this scheme. | - The proposed method adopted two hybrid approaches<br>a- Statistical analysis of the network traffic and time series to find out statistics parameters that result from DDoS SYN flood attack.<br>b- Determine the intensity of DDoS SYN flood attack by fuzzy logic.<br>The hybrid approach is accurately detect the DDoS SYN flood attack. |
| [43] | Multi-Layer Perceptron (MLP) neural network is used to detect DDoS flood attack. | - False positive (FP) and False negative values are generated in classification phase of artificial neural network.<br>- Irrelevant neural network output is generated, which can consider as False Negative (FN) values.<br>- The long training time of the neural network was mostly due to the huge number of training vectors of computation facilities. | - This technique is slightly better in classification results compared with other methods, where this technique is more computationally and memory wise efficient.<br>- This technique concludes that there is more to do with the irrelevant output in the field of artificial neural network based security systems. |

## 7.4 Other Detection Schemes

There are many other detection mechanisms to detect denial of Service (DoS) apart from those mentioned above.

Data mining techniques [44]have been adopted for detection of DoS attacks in order to discover useful information and hidden relationships in large data repositories[45][46]. In[46], the network traffic and network packet protocol status model was extracted based on FCM cluster algorithms and Apriori association. The proposed system was viable in Local Area Network (LAN).[47] adopted simple proactive anti-DDoS framework. Initially DDoS attack features were analyzed. Then, variables based on these were selected by investigation many procedures. Finally, the k-nearest neighbour (K-NN),was applied to classify and detect DDoS SYN attacks for each phase. The results show that this method can classify DDoS phases correctly.

## 8. PERFORMANCE MEASURES

In conclusion, previous researches adopted various mechanisms to detect the Denial of Service (DoS) attacks such as those based on the router data structure, statistical analysis, neural network and fuzzy logic which have their respective advantages and weakness. Nevertheless, the detection of the DoS attack is still complex. When a user sends many packets with a spoofed IP addresses, it will be difficult to model the traffic flow according to the mechanisms above as the packet flow will be in bursty (non-linear) during the attacking period. On the other hand, some users send this attack under the normal flow rate which causes difficulty in accurately diagnosing the traffic flow and detecting a DoS attack due to false positive and false negative values. As a result, a large number of pointless requests is sent, leading to the shutdown of available services.

The performance of Denial of Service (DoS) detection schemes may be measured in terms of CPU time, memory consumption, false positive and false negative detection and accuracy of detection at high rates and low rates. The performance comparison is summarized in Table 5.

Table 5.Comparison of the performance measures of each category

| Category | Papers | CPU Time | Memory Consumption | False Positive (FP) | False Negative | Accuracy Detection | |
|---|---|---|---|---|---|---|---|
| | | | | | | Detection high Rate traffic | Detection Low rate traffic |
| **Router based data structure** | [22] | flexible | flexible | high | NA | very good | no |
| | [23] | flexible | flexible | high | NA | good | good |
| | [[24] | flexible | flexible | high | NA | very good | fair |
| | [23] | low | low | high | fair | very good | no |
| | [24] | flexible | flexible | high | NA | very good | fair |
| **Statistical Analysis** | [29] | high | high | high | NA | good | no |
| | [30] | high | high | high | high | good | no |
| | [31] | flexible | flexible | high | NA | good | no |
| | [32] | NA | NA | high | low | good | no |
| | [33] | high | high | high | NA | good | no |
| | [44] | high | high | high | NA | very good | NA |
| | [45] | high | high | NA | NA | good | NA |
| **Artificial Intellige** | [40] | high | NA | NA | high | very good | NA |
| | [38] | flexible | flexible | high | high | good | no |
| | [39] | high | high | flexibl | NA | good | no |

| nce | | | | e | | | |
|---|---|---|---|---|---|---|---|
| | [41] | high | high | NA | NA | very good | no |
| | [42] | high | high | flexible | high | good | no |
| | [43] | flexible | low | high | high | very good | no |

## 9. CONCLUSION

In this new era of mobile computing, organizations adopt various strategies and applications to ensure the security of precious information. Intrusion detection has become an integral part of the information security process. This paper presents a summary of the state-of-the-art of detection schemes for SYN-Flooding attacks. The detection schemes for SYN- Flooding attacks have been classified broadly into three categories – detection schemes based on the router data structure, detection schemes based on statistical analysis of the packet flow and detection schemes based on artificial intelligence. The advantages and disadvantages for various detection schemes under each category have been critically examined. The performance measures of the categories have also been compared.

**Authors**

**Mehdi EbadyManaa** is currentlypursuing his Master of Science  at University Utara Malaysia. He will graduate in 2012. His Bachelor of Science is from University of Babylon in Iraq.  His main area of research is network security. He is currently focusingon the detection of SYN-flooding attacks in Local Area Networks.

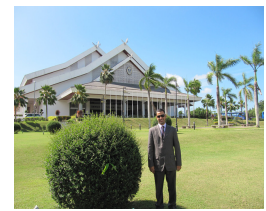

**Dr. Angela Amphawan**received her Ph.D. degree from University of Oxford, United Kingdom. In her doctoral research, she successfully increased the bandwidth of a multimode fiber channel using a novel holographic selective launch.  Her research interests include radio and optical communication technologies.